# Hardware Density Reduction To Avoid Proximal Junction Failure In Adult Spine Surgery: In Silico Case Studies and Virtual Cohort

**Morteza Rasouligandomani[1*], Alex del Arco[2], Tomaso Villa[3], Luigi La Barbera[3], Ferran Pellisé[4], Miguel González Ballester[1,5], Fabio Galbusera[6], Jérôme Noailly[1]**

[1]BCN MedTech, DTIC, University Pompeu Fabra, Barcelona, Spain

[2]Hospital del Mar, IMIM, Barcelona, Spain

[3] Polytechnique University of Milan, Italy

[4]University Hospital Vall d'Hebron, Barcelona, Spain

[5]ICREA, Barcelona, Spain

[6]Schulthess Klinik, Zürich, Switzerland

**\* Correspondence:**
Corresponding Author: Morteza Rasouligandomani, email: morteza.rasouli@upf.edu

**Abstract**

*Background*
Proximal Junctional Failure (PJF) is a post-operative complication in adult spine surgery, often requiring reoperation. Osteotomy is often used in revision surgeries, leading to 34.8% complications. Hence, suboptimal decisions might be extending hardware without osteotomy, which yields to severe Global Alignment and Proportion (GAP) scores. High GAPs increase PJF risk, but Hardware Density Reduction (HDR) might limit it.

*Methods*
Two clinical cases were evaluated: 1) Initially operated with hardware extended to T10, GAP 10; 2) PJF at T11 and hardware extended to T3, GAP 11. Two patient-personalized spine FE models were constructed through Statistical Shape Modelling (SSM) and mesh morphing. Intervertebral Disk (IVD) fiber strain, screw pull-out force, and rod stress were evaluated for the cases 1) and 2), also for 91 virtual HDR scenarios with different GAP scores, using Finite Element (FE) simulations. Different rod and bone material properties were also assessed.

*Results*
HDR could decrease IVD fiber strain (-70% at most) and increase screw pull-out forces (+142% at most) for cases with Ti rod and normal bone. Cr-Co rod and osteopenia, and osteoporotic bones had high PJF risk. Trade-off analyses could determine the best configurations avoiding PJF. Virtual cohort study showed that GAP 12 and 13 could not avoid PJF in any HDR scenarios either with Ti or Cr-Co rods. HDR in a UIV T10 virtual patient with GAP 11 could not de-risk in case of Cr-Co rods. UIV T3 with GAP 13 could not benefit any HDR strategy, independently of rod properties. In contrast, Ti rods might allow HDR to de-risk GAP 12 patients with UIV T3.

*Conclusions*
HDR could avoid PJF in the patients with medium high GAP scores, depending on the screw reduction pattern, and bone and rod material properties. Remarkably, HDR technique might avoid excessive spine surgeries and minimize the surgery cost.

## 1. Introduction

Adult Spine Deformity (ASD) is a multifactorial condition that consists in the emergence of abnormal lumbar or thoracolumbar sagittal balance in adults. Conservative and surgical treatments are decided, based on Health-Related Quality of Life (HRQOL) (Yin et al., 2016). Surgical treatments aim to restore spine alignment and relieve pain. 8.7% to 36% of spine surgeries may have mechanical complications post-operatively (Akıntürk et al., 2022). Proximal Junctional Failure (PJF) is an early (first 3-6 months) post-operative mechanical complication observed in at least 8.7% of sagittal imbalance spine surgery (Acaroglu et al., 2017). PJF is classified as: adjacent Intervertebral Disk (IVD) degeneration, ligament disruption, hardware loosening, and fractures at the Upper Instrumented Vertebrae (UIV) or in the adjacent vertebrae (Hart et al., 2013).

Patients with PJF are reoperated (revision surgery) because of mechanical instability, pain or neurological deficit, and dramatic Proximal Junction Angle (PJA) (S.-J. Park et al., 2021). Since spine re-alignment is one of the goals of revision surgery due to PJF, orthopedic surgeons typically remove the previous instrumentations, an recorrect the spine curvature using both lumbar osteotomy and a new hardware configuration extended to the upper levels (Kim et al., 2009; Nguyen et al., 2016). Because lumbar osteotomy induces 34.8% of post-op complications in elderly patients, it is not optimal to recorrect lumbar curvature of patients with PJF (Berjano & Aebi, 2015). Hence, the suboptimal decision to simply extend the hardware to the upper levels without revising the prior instrumentations might be made.

Spine sagittal misalignment due to previous PJF increases the Global Alignment and Proportion (GAP) score (Yilgor et al., 2017). GAP predicts post-op mechanical complications based on the proportional radiographic parameters. Rasouligandomani et al. found that a GAP threshold of 7 i.e., GAP values upper than 7, reflects a strong risk of PJF (Rasouligandomani et al., 2023). Therefore, patients with high GAP scores (7 to 13) shall suffer a 2nd PJF after the revision surgery. However, some prevention techniques can be used during the second operation, to limit the risk of 2nd PJF and third surgery, even with high GAP values. Remarkably, Cummins et al. showed that the excess of mortality increased with spine revision surgeries (Cummins et al., 2022). Hence, early PJF prevention techniques are critically needed for patients with high GAP values even after the first surgery.

PJF prevention techniques can be classified as: Hardware Density Reduction (HDR) in the proximal levels of instrumentations; sublaminar tethers; sublaminar tape; pretensioned suture loops; transverse and laminar hooks (Doodkorte et al., 2021). In the last decade, a variety of prevention (topping-off) techniques have been biomechanically assessed (Doodkorte et al., 2021). Yagi et al. (Yagi et al., 2020) tested several posterior tether scenarios and by using Finite Element (FE) simulations, they found that tethers may decrease Proximal Junctional Kyphosis (PJK) development. Bess et al. also found that posterior tethers could create a more gradual transition in adjacent-segment stress, which it limits the biomechanical risk factor for PJK (Bess et al., 2017). Metzger et al. reported that supralaminar hooks at the top of a multilevel posterior fusion construct reduces the stress at the proximal un-instrumented motion segment, which reduces risk of PJK (Metzger et al., 2016). Polycarbonate-urethane rod could also decrease the risk of screw pull-out (Jacobs et al., 2017). The objective of all these techniques is to reduce the risk of proximal stress concentrations. Accordingly, a few studies also focused on the screw density reduction. In a multivariable analysis, Durand et al (Durand et al., 2022) studied the correlation between the number of screws and the risk of PJF, and they found that patients with less than 1.8 screws per level had lower risk of PJF. Although, clinical practice showed that screw density reduction could decrease PJF risk, it is still unclear whether optimal screw density patterns can be anticipated to prevent PJF in the revision of ASD correction surgery.

Recently, Finite Element (FE) simulations could help orthopedic surgeons to explore the biomechanical response of the spine in detail. Zhu et al (Zhu et al., 2019) found that Upper Instrumented Vertebra (UIV) at proximal or distal thoracic vertebras may increase risk of PJF, in case of long instrumentations. However, there is lack of computational studies to investigate biomechanical responses to screw density reduction. Accordingly, this chapter focuses on simulating and quantifying the biomechanical effects of HDR as one of the most straight forward PJF prevention techniques. First, it aims to explore biomechanical responses of screw density reduction to avoid $1^{st}$ and $2^{nd}$ PJF for the patients who have severe GAP score using FE simulations. Second, it explores best screw density configurations through a trade-off analysis of different biomechanical descriptors. Third, it simulates the interactions between screw density configurations and PJF prevention with different GAP scores.

## 2. Material and methods
### 2.1. Clinical cases
#### 2.1.1. Case 1 (T10-int)

Two clinical cases were chosen, to study HDR effects through FE simulations. The first case (female, 55 y.o., anonymized) was extracted from a retrospective observational cohort with 57 PJF cases, collected at IRCCS (Istituto Ortopedico Galeazzi, Milan) database (Rasouligandomani et al., 2023). It fulfilled the following criteria:

- PJF at first operation
- Hardware extension to the lower thoracic levels
- No PJF prevention techniques in $1^{st}$ PJF at UIV/UIV+1,2
- GAP score 9 for $1^{st}$ operation
- Age 50 to 75 years
- PJF indications: Hart criteria (R. A. Hart et al., 2013)

This first case was named T10-int. PJF occurred at the T10 level after 8 months post-operation (Figure 1). Sagittal spine geometrical parameters were calculated through the sterEOS software (EOS Imaging, Paris, France) (Rasouligandomani et al., 2023), and were confirmed by an expert spine orthopedic surgeon, and the case is explained as follow: Ang et al. showed that reciprocal post-operative increase in Thoracic Kyphosis (TK) correlates with an increase in the spinopelvic mismatch (Pelvic Incidence (PI) – Lumbar Lordosis (LL)) for lumbar fusions (Ang et al., 2021). In other words, LL increase may lead to reciprocal increase in TK. Since TK was initially high, 60º (normal TK 35º to 50º (Abrisham et al., 2020)) for this patient, postoperative LL increase made TK worser, leading to definite PJK/PJF. Therefore, LL was increased by only 7º postoperatively. Misaligned LL led to a GAP score of 10 because of severe misalignment in the first operation, which stands for a high risk of mechanical complications like PJF.

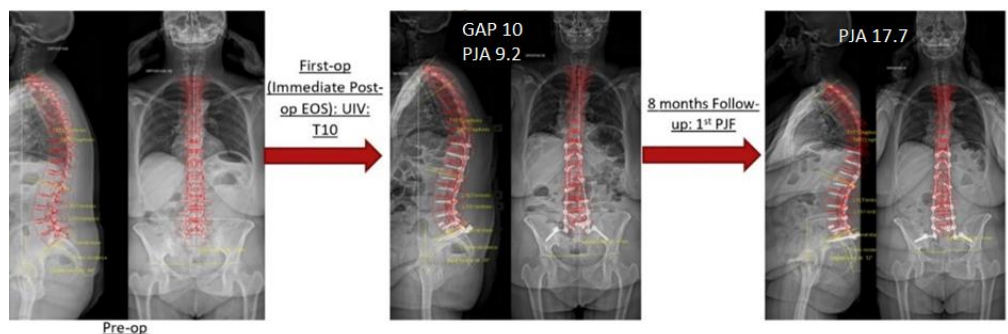

*Figure 1: Pre-op, first-op and first follow-up for the first case (T10-in): female, 55 y.o., PJF at T10 (PJA: Proximal Junction Angle; GAP: Global Alignment and Proportion)*

### 2.1.2. Case 2 (T11-T3)

The second case (female, 76 y.o., anonymized) was chosen out of the IMIM (Hospital del Mar, Barcelona) database, and fulfilled the following inclusion criteria:

- At least 2 successive PJF
- Hardware extension, from initial instrumentation at $1^{st}$ PJF in the lower thoracic levels, to the upper thoracic levels, before $2^{nd}$ PJF
- No spine curve correction between $1^{st}$ and $2^{nd}$ PJF
- No PJF prevention techniques before $2^{nd}$ PJF at UIV/UIV+1,2
- GAP score 9 for $2^{nd}$ operation
- Age 50 to 75 years
- PJF indications: Hart criteria (R. A. Hart et al., 2013)

This second case was named T11-T3. The first PJF occurred at T11. Then, instrumentation was extended from T11 to T3, while preserving the previously existing hardware. A second PJF occurred at T3 level 8 months after the $1^{st}$ revision surgery (Figure 2). MRI and X-rays images were available for this case, and sagittal spine geometrical parameters were calculated using the Surgimap software (Surgimap®, Massachusetts, USA). They were confirmed by an orthopedic surgeon. A GAP score of 11 was calculated after the $2^{nd}$ operation, reflecting a severe misalignment (Yilgor et al., 2017). Spine curvature was not corrected during this $2^{nd}$ operation.

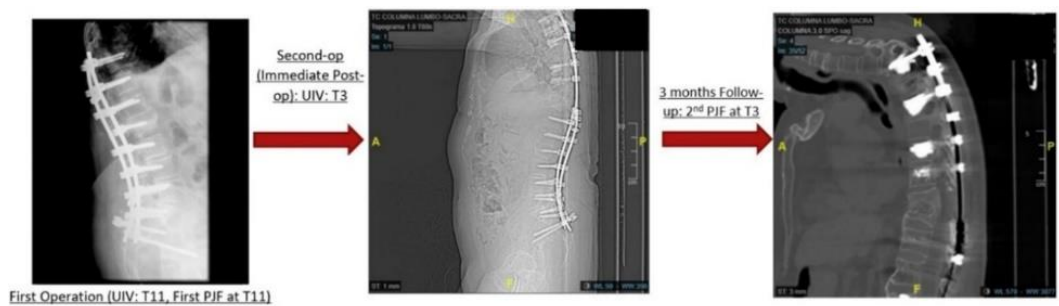

*Figure 2: First-op, second-op and follow-up for case T11-T3: female, 76 y.o., 1st PJF at T11 and 2nd PJF at T3.*

Sagittal spine geometrical parameters for the T10-int and the T11-T3 cases were used later for in-silico patient-personalized modelling.

### 2.2. In-silico patient-personalized modelling

Two patient-personalized thoracolumbar FE models were generated based on the geometrical parameters of the pre-op and 2$^{nd}$ surgery, for the 1$^{st}$ case (T10-int) and 2nd case (T11-T3), respectively. Patient-personalized FE model reconstruction pipeline is shown in Figure 3.

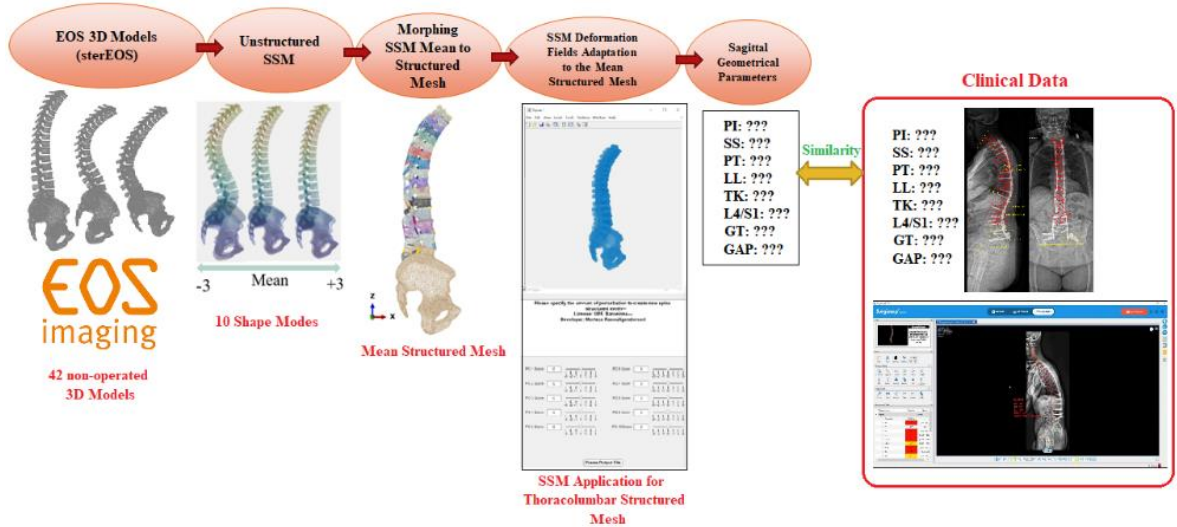

*Figure 3: Pipeline to generate patient-personalized thoracolumbar spine models.*

Overall, EOS images of 42 pre-operated patients were gathered. Inclusion criteria were age 50-75; Pelvic inclination>20°; SVA>5cm. Images were converted into 3D surface models by using the sterEOS software, and they were later aligned by rigid registration. Principal Component Analysis (PCA) was used to explore the shape modes representing the shape variabilities and a mean model, and to extract the principal modes of variation by computing the eigenvectors of the covariance matrix (Jolliffe & Cadima, 2016). 10 main shape modes were retained to represent 95% of the cumulative variance of patient spine morphologies. Shape mode combinations was then used to generate new shapes and obtain models that meet the geometrical parameters of the patient spines of the T10-in and T11-T3 clinical cases:

$$\text{New Shape} = \text{Mean} + Pb \tag{1}$$

where Mean was the average shape, P was matrix of eigenvectors of the covariance matrix, and b was a user-defined vector of weights.

### 2.2.1. Thoracolumbar FE mesh generation

The unstructured 3D surface mean model was subsequently transformed into a structured hexahedral FE model through the morphing of a pre-existing 3D FE structural mesh template of the osteo-ligamentous spine (Malandrino et al., 2015) to the geometry of the mean shape model. To adapt the 3D structural meshes of the soft tissues along with the ones of the vertebra, a Bayesian Coherent Point Drift (BCPD++) (Hirose, 2021) algorithm was used to drift coherently the point clouds of the osteo-ligamentous spine mesh template, based on a Gaussian Mixture Model (GMM) in a Bayesian form. Overall, the surface meshes of thoracolumbar vertebras were mapped to the hexahedral meshes, and 17 IVD meshes were interpolated between adjacent vertebras. Euclidian distance between morphed FE, and unstructured mean model was evaluated to increase resemblance between two models. Six groups of ligaments were included as truss elements: anterior and posterior longitudinal ligaments; capsular ligaments, ligamentum flavum, the supraspinatus and interspinous ligaments, according to pre-existing landmarks defined on the vertebra of the 3D mesh FE template (Malandrino et al., 2015).

### 2.2.2. Patient-personalized FE models

The unstructured statistical and the structured FE thoracolumbar mean models were aligned rigidly. The closest points between two mean models were found by using an Iterative Closest Point (ICP) algorithm. Then, the unstructured SSM covariance matrix was assigned to the points of the FE mesh, leading to an automatized SSM able to reproduce osteo-ligamentous spine hexahedral FE mesh.

16807 structured patient-personalized models were further sampled by combining the first 5 shape modes of the SSM, in which each shape mode was discretized into 7 Standard Deviations (SD). Samples data was further shared in Zenodo data repository (https://doi.org/10.5281/zenodo.8107354). All sagittal and frontal geometrical parameters were also calculated for the 16807 FE models, and the geometrical parameters of the pre-op and the 2nd surgery, for the T10-in and the T11-T3 case, respectively, were searched among the shared database to select two patient-personalized FE models which had highest similarity (Figure 4a for 1$^{st}$ case, and Figure 4b for 2$^{nd}$ case). Clinical spinopelvic parameters for the T10-in and the T11-T3 cases were compared with generated models in Table 1.

*Table 1: Clinical and model spinopelvic parameters for the T10-in and the T11-T3 cases*

| *Cases* | *Measurements* | *PI* | *SS* | *LL* | *GAP* | *Max Difference %* |
|---|---|---|---|---|---|---|
| pre-op T10-int | Model | 69 | 28 | -41 | 10 | 2.85% |
| | Clinical | 71 | 28 | -41 | 10 | |
| 2$^{nd}$-op T11-T3 | Model | 37.29 | 24 | -32 | 11 | 5.43% |
| | Clinical | 37.15 | 22.73 | -31.16 | 11 | |

Results in Table 1 showed that generated models had an acceptable range of differences compared to the clinical measurements.

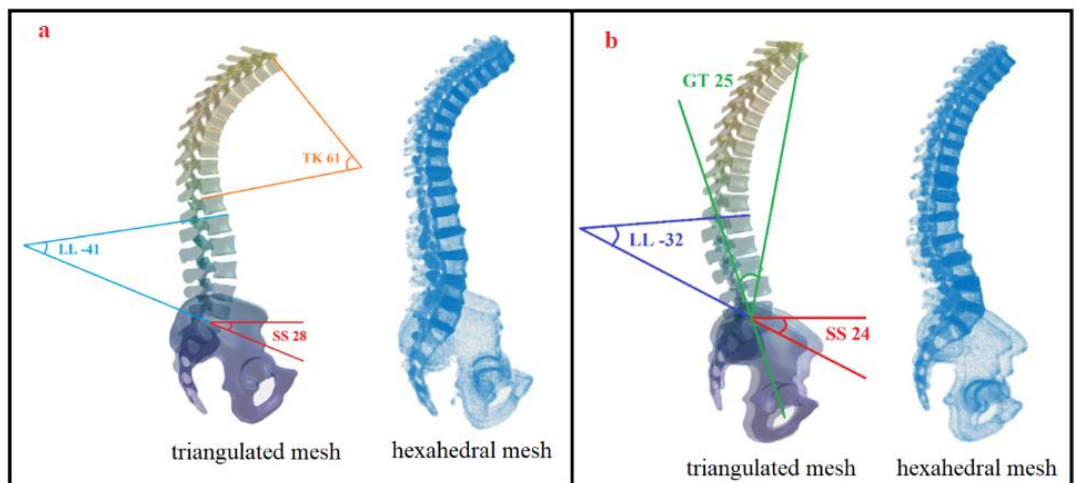

*Figure 4: (a) Triangulated mesh and osteo-ligamentous spine hexahedral FE mesh for case 1 (T10-int, pre-op); (b) Triangulated mesh and osteo-ligamentous spine hexahedral FE mesh for case 2 (T11-T3) (NOTE: to compare the models visually, ligament elements were removed from the FE meshes).*

### 2.3. Instrumentations

The screws were designed through Computer-Aided Design (CAD) software. Starting from a real screw (showed in Figure 5a), a simplified model (Figure 5b) was created, ignoring the threads, and combining small components in 2 main parts: the head and shaft of the screw.

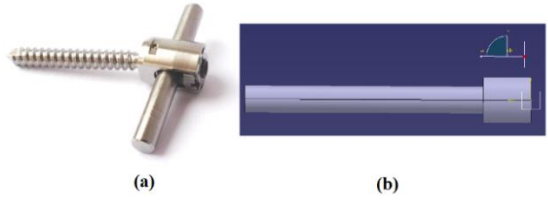

*Figure 5: (a) Poly-axial real screw; and (b) simplified screw created by a CAD software.*

According to manufacturer standards, generic screw dimensions were introduced for different spine zones (Table 2).

*Table 2: Screw dimensions*

| Zone | Sacrum | Lumbar | T12 to T8 | T7 to T1 |
|---|---|---|---|---|
| **Dimensions (Length×Diameter) in mm** | **35×7** | **45×6** | **40× 5** | **35×4.5** |

The rods models were further implemented as beam element, and following smoothed screw head trajectories, and the trajectory line was radially expanded to reach a standard 5.5 mm diameter. Moreover, preparation holes were made through pedicle and the vertebral body of pre-op model by using Boolean/cut operations in Abaqus to place the screws. Therefore, screw shafts were relocated into these preparation holes through the kinematical constraints. Xu et al., showed that immediately after the pedicle screw-instrument surgery, the screw and bone are not fully bonded where the relative movement and friction force between the screw and surrounding bony tissues should be considered as contact connection (Xu et al., 2019). In this study, the screw-bone interface is described surface-to-surface frictional contact with friction coefficient of 0.61 (Çetin & Bircan, 2022). Varghese et al., demonstrated that screw pull out force is the resultant of axial and perpendicular forces acting on the screw (Varghese et al., 2017). Hence, average contact force (normal + shear) was calculated in this study, as the screw pull-out force. Half-thread angle, pitch distance, bolt major and mean thread diameter for pedicle screws were defined as clearance feature. Tie constraints were also defined to ensure a fixed connection between the screw head and shaft (body). Tie contact was also used to couple the tetrahedral mesh of the pedicles with the hexahedral mesh of the vertebral bodies.

### 2.3.1. T10-int case

The T10-int case model was virtually instrumented, to represent the 1st operation scenario. A virtual dummy model of the first operation was generated as well. This dummy model was matched with operation image in Figure 1: model parameters: PI: 69, SS: 32, LL: -48, GAP: 10, vs. clinical data: PI: 70, SS: 30, LL: -48, GAP: 10. The first-op dummy model was subsequently instrumented using screws and rods, matching the bi-planar EOS radiographs presented in Figure 1. First-op dummy model with virtual pedicle screw insertion for the T10-int case is shown in Figure 6a.

To simulate the residual stresses induced by the surgical correction, the T10-int pre-op patient-personalized thoracolumbar FE model, and the virtual instrumentation defined with the first-op dummy model were simultaneously imported in Abaqus v2019 (SIMULIA). The alignment of the Pre-op patient-personalized model was corrected through FE simulation thanks to imposed kinematic constraint that aimed to reach a proper positioning of the vertebra with respect to the first-op instrumentation and match the clinical data in Figure 1.

### 2.3.2. T11-T3 case

The T11-T3 second-op model was instrumented virtually without correcting the spine curvature (GAP 11) compared to first-op. In other words, hardware was just extended from T11 to T3 without spine curvature correction, and both first-op and second-op had the same deformity (GAP 11). Therefore, second-op model (shown in Figure 4b) was subsequently instrumented using similar screw and rod modelling as described for the T10-int case, according to the radiographs presented in Figure 2. The T11-T3 second-op model with virtual instrumentations (CAD software) is shown in Figure 6b.

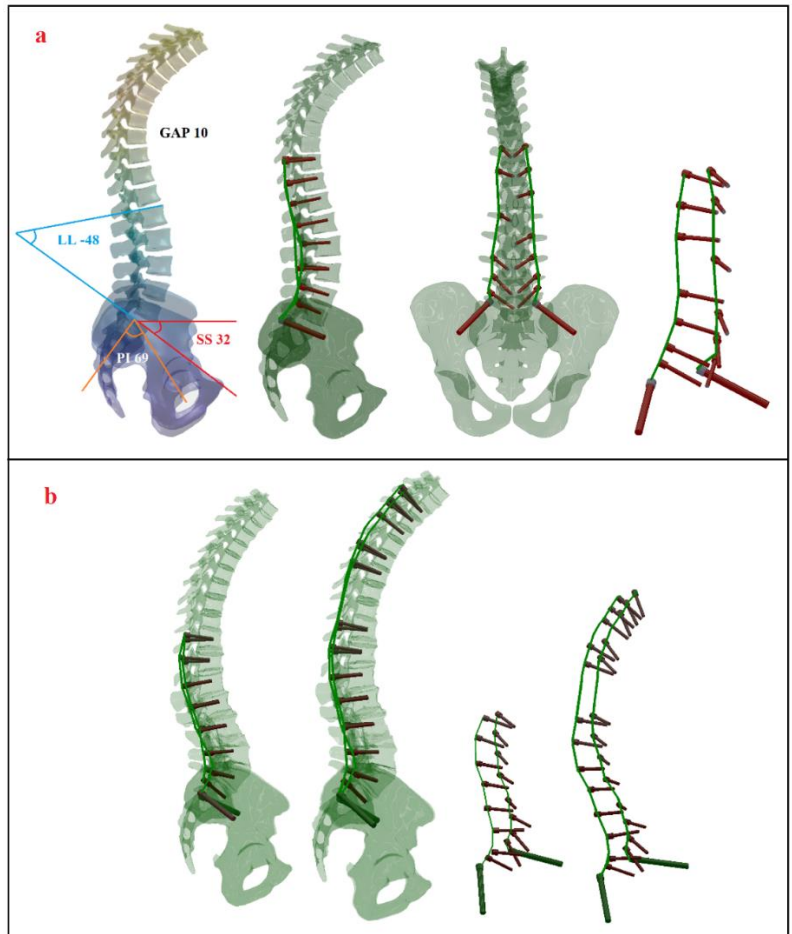

*Figure 6: (a) First-op dummy model generation for 1st case and the dummy instrumented model in a CAD software; (b) Second-op instrumented model for 2nd case in a CAD software.*

### 2.4. Hardware density scenarios

The HDR scenario to be implemented in silico was defined according to clinical data. Literature review showed that Screws per Level (SpL, i.e., total number of screws divided by the number of levels) <1.8 decreases risk of PJF (Durand et al., 2022). Among 57 control cases (without any post-op complications) from a previously gather ASD patient cohort surgically treated at IRCCS (Rasouligandomani et al., 2023), 23 cases (40%) had at least one level with missing screws (yellow cells, Table 3), and for all the 23 cases, SpL was lower than 1.8 (Table 3).

*Table 3: 23 selected control cases with HDR (green cell means fused level, yellow cell means non-instrumented level)*

| Case No. | Level of Fixation | GAP score | L5 | L4 | L3 | L2 | L1 | T12 | T11 | T10 | T9 | T8 | T7 | T6 | T5 | T4 | T3 | Number of screws/ Levels |
|---|---|---|---|---|---|---|---|---|---|---|---|---|---|---|---|---|---|---|
| 1 | T3 | 0 | G | G | G | G | G | G | Y | Y | G | G | Y | G | Y | G | G | 1,33 |
| 2 | T3 | 0 | G | G | G | G | Y | G | Y | G | Y | G | G | G | Y | Y | G | 1,2 |
| 3 | T11 | 1 | G | G | G | Y | G | G | G | | | | | | | | | 1,71 |
| 4 | T11 | 1 | G | G | G | Y | G | G | G | | | | | | | | | 1,71 |
| 5 | T11 | 1 | G | G | G | Y | G | G | G | | | | | | | | | 1,71 |
| 6 | T3 | 1 | G | G | G | G | G | Y | Y | Y | G | G | Y | G | G | G | G | 1,33 |
| 7 | T4 | 2 | G | G | G | G | Y | G | G | Y | G | G | Y | G | G | G | | 1,71 |
| 8 | T10 | 2 | G | G | Y | G | G | G | G | G | | | | | | | | 1,75 |
| 9 | T10 | 3 | G | G | G | Y | G | G | Y | G | | | | | | | | 1,5 |
| 10 | T4 | 3 | G | G | G | G | Y | G | Y | G | Y | G | Y | G | Y | G | | 1,28 |
| 11 | T10 | 3 | G | Y | G | G | G | G | G | G | | | | | | | | 1,5 |
| 12 | T11 | 3 | G | Y | G | Y | G | G | G | | | | | | | | | 1,42 |
| 13 | T4 | 5 | G | G | G | Y | G | Y | G | Y | G | G | G | G | G | G | | 1,57 |
| 14 | T4 | 5 | G | Y | G | Y | G | Y | G | Y | G | Y | Y | G | G | G | | 1,14 |
| 15 | T3 | 5 | G | G | G | Y | G | G | G | G | G | G | Y | G | Y | G | G | 1,6 |
| 16 | T9 | 5 | G | G | G | Y | G | G | Y | G | G | | | | | | | 1,55 |
| 17 | T9 | 5 | G | G | G | G | G | G | G | G | G | | | | | | | 1,77 |
| 18 | T4 | 6 | G | G | Y | G | G | G | G | G | G | Y | G | G | G | G | | 1,28 |
| 19 | T10 | 6 | G | Y | G | G | Y | G | Y | G | | | | | | | | 1,5 |
| 20 | T10 | 8 | G | G | G | G | Y | G | Y | G | | | | | | | | 1,5 |
| 21 | T4 | 8 | G | G | G | Y | G | G | G | G | Y | G | Y | G | G | G | | 1,42 |
| 22 | T10 | 9 | G | Y | Y | G | Y | G | Y | G | | | | | | | | 1 |
| 23 | T3 | 9 | G | G | Y | G | G | G | Y | G | Y | G | Y | G | Y | G | G | 1,46 |

For the HDR in the T10-int case model, 6 cases were identified in Table 3, with UIV at T10. Different screw missing scenarios were observed between UIV-1 to UIV-4 as topping-off techniques. Accordingly, 5 virtual HDR scenarios plus the real case were chosen for 1st case (Figure 7a). For the HDR in the T11-T3 case model, 5 cases had UIV at T3 in Table 3, with different screw missing scenarios between T11 and T3. Accordingly, 6 virtual HDR scenarios plus the real case were chosen for 2nd case (Figure 7b).

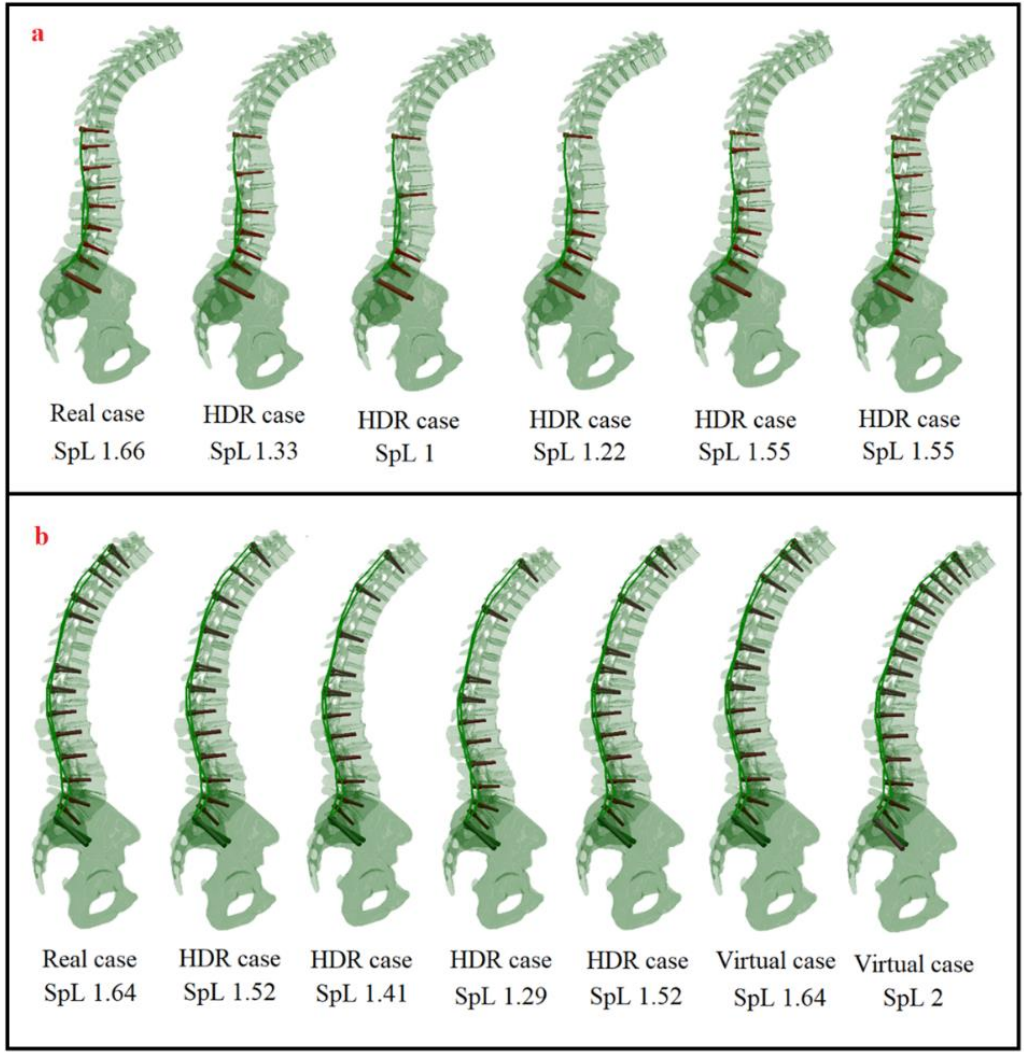

*Figure 7: Different instrumentation scenarios for 1st (a), and 2nd (b) cases*

**2.5. In-silico models for HRD effects with different GAP scores**

Out of the T10-int and the T11-T3 clinical case models, further models were generated to explore the impact of the GAP score on the biomechanical performance of the HDR. It shall allow medical doctors to select the best HDR scenario based on the GAP score and biomechanical rationales. 91 in-silico models with GAP scores from 7 to 13 (PJF GAP threshold 7 (Rasouligandomani et al., 2023)) and different HDR scenarios for UIV at T10 and T3 were virtually implemented thanks to the virtual cohort shared in Zenodo (Section 2.2). 42 in-silico models for UIV at T10 and 49 in-silico models for UIV at T3 were generated. We assumed that the simulated cases correspond to suboptimal cases when no further correction is performed after a PJF, justifying HDR strategies. Hence, no kinematic constraints to correct the spine curvature were introduced. Subsets of 7 models for UIV at T10 and 7 models for UIV at T3 with SpL 2 (HD 100%) are presented in Figure 8, among the 91 in-silico models. For each subset, two HDR examples are represented as well: 5 scenarios for UIV at T10 with GAP 10 (SpL 1.5, 1, 1.25, and 1.75(2)), and 6 scenarios for UIV at T3 with GAP 11 (SpL 1.64(2), 1.52(2), 1.29, and 1.41) are also presented (Figure 8). The rest of in-silico models had the HDR topology the same as the HDR configuration in Figure 8, and the only difference among models was the spine deformity and GAP score.

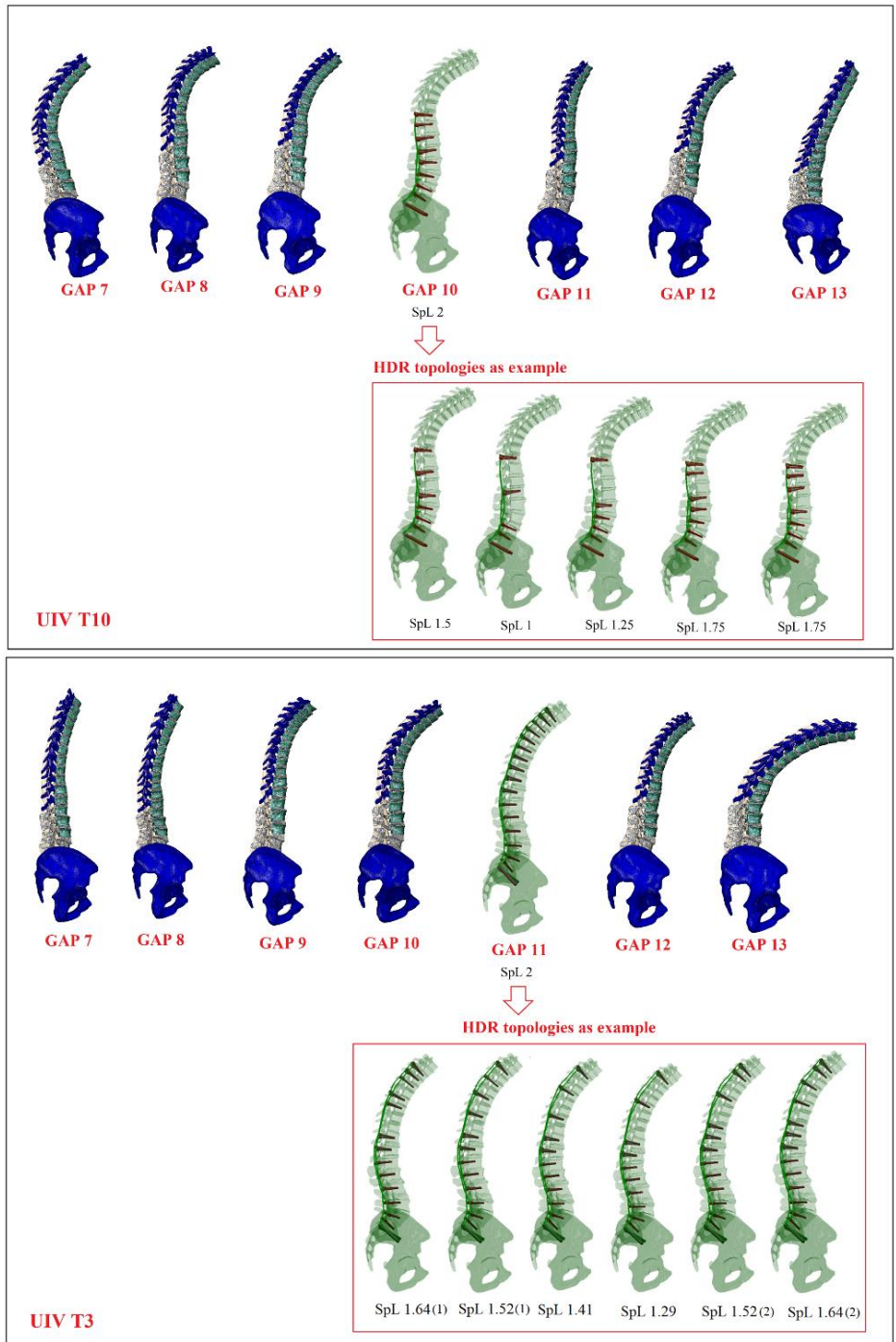

*Figure 8: In-silico models for GAP 7 to 13 with different HDR scenarios.*

### 2.6. Material properties

Two standard materials were considered to simulate the screws and the rods. Titanium alloy (Ti6Al4V) is used with the following material properties: Young Modulus (E) 115 GPa; poison's ratio ($\upsilon$) 0.3; Yield Stress ($\sigma_y$) 880 MPa (Metals Handbook, 1961). On the other hand, a Cobalt-Chromium (Cr-Co) instrumentation was modelled with the following material properties: E = 210 GPa; $\upsilon$ = 0.29; $\sigma_y$ = 1585 MPa (Metals Handbook, 1961).

Three material properties of bone, i.e., normal, osteopenia, and osteoporotic were used for the vertebrae (Table 4). Generic properties from the literature were used for each bone quality since patient-specific bone mineral density was not available in EOS images. We assumed that 14 screw threads were geared into the bone (fully inserted screw). Hence, independent ultimate screw pull-out forces for normal and osteoporotic bones were 2300N, and 900N respectively, as calculated by Jendoubi et al (Jendoubi et al., 2018). Ultimate screw pull-out force for osteopenia bone was considered halfway between normal and osteoporotic bone.

*Table 4: Vertebra material properties*

| | E (MPa) | υ | References |
|---|---|---|---|
| **Normal bone density** | | | |
| Cortical bone | 12000 | 0.3 | (Cowin SC, 1991) |
| Trabecular bone | 200 | 0.315 | (Lu, 1996) |
| Pedicle bone | 3500 | 0.25 | Calculated by (Shirazi-Adl et al., 1986) |
| **Osteopenia bone** | | | |
| Cortical bone | 8350 | 0.3 | Assumed as mean properties between normal and osteoporotic bone |
| Trabecular bone | 173 | 0.315 | Assumed as mean properties between normal and osteoporotic bone |
| Pedicle bone | 3150 | 0.25 | Assumed as mean properties between normal and osteoporotic bone |
| **Osteoporotic bone** | | | |
| Cortical bone | 4700 | 0.3 | (Kasra et al., 1997) |
| Trabecular bone | 146 | 0.315 | (Jensen et al., 1990) |
| Pedicle bone | 2800 | 0.25 | Assumed by (Galbusera, Bellini, Anasetti, et al., 2011) |

The Young´s modulus of the osteoporotic pedicle elements assumed as 80% reduction compared to the normal posterior elements (Galbusera, Bellini, Anasetti, et al., 2011). Moreover, in absence of any specific reference for osteopenia bone, an average Young Modulus between normal and osteoporotic bone stiffness was assumed (Table 4).

Ligaments were modelled as hypoelastic materials (Noailly et al., 2012). Moreover, we assumed that before starting the surgery, IVDs were pre-swelled. The swelling process of IVDs can be influenced by osmotic pressure. The nucleus pulposus, being a gel-like structure, contains proteoglycans and water. Proteoglycans have negatively charged sulfated groups, which attract positively charged ions (cations) and water through electrostatic interactions. This results in an osmotic gradient within the disc. In addition, IVD material consists of two phases: (1) A solid phase made up of structural macromolecules (collagen, elastin, and proteoglycans) as well as cells, and (2) a fluid phase that consists of water and solutes, leading to consider IVDs as poroelastic materials (Malandrino et al., 2015). Therefore, IVDs were modelled as osmo-poroelastic materials. Annulus fibrosus was modelled according to the anisotropic hyperelastic formulation. The annulus fibrosus, like many others soft biological materials and tissues, has an anisotropic behavior due to the presence in the microstructure of fibers, acting along different preferential directions. The Holzapfel-Gasser-Ogden formulation was chosen to model this structure. The form of the strain energy potential implemented is (Holzapfel et al., 2005):

$$U = C_{10}(\bar{I}_1 - 3) + \frac{1}{D}\left(\frac{J_{el}^2 - 1}{2} - \ln J_{el}\right) + \frac{K_1}{2K_2} \sum_{\alpha=1}^{N} (\exp[K_2 \langle \bar{E}_\alpha \rangle^2] - 1)$$

$$\bar{E}_\alpha \stackrel{def}{=} k(\bar{I}_1 - 3) + (1 - 3k)(\bar{I}_{4(\alpha\alpha)} - 1)$$

(2)

where $K_1$ is fiber stiffness, $K_2$ is fiber nonlinearity, k is fiber dispersion, $C_{10}$ is constant related to the stiffness of the matrix, N is number of families of fibers and D is incompressibility modulus. For the annulus fibrosus two families of fibers, were oriented with an angle of $\pm 30$ with respect to the local horizontal plane. The values adopted for the parameter of the strain energy function were found in the literature: $K_1$=2.02 MPa, $K_2$=86.53, k=0.113, $C_{10}$=0.0154 MPa and D=0.306 1/MPa (Malandrino et al., 2015).

### 2.7. Loading conditions

Body Mass (BM) distributions were translated into punctual static loads in function of the contribution expected from the rest of the upper body, as detailed elsewhere (Toumanidou & Noailly, 2015). The resulting loads (F) were proven to stand for excellent biomechanical descriptors, for the discrimination of PJF (Rasouligandomani et al., 2023). Briefly, these loads are positioned along an eccentric path designed to pass through the center of mass of each body level, at an effective distance ($d_{eff}$) from the Vertebral Center (VC) position (Toumanidou & Noailly, 2015). So, F was calculated as the sum of the upper body loads at each vertebrae level, and $d_{eff}$ was calculated through the Huygens-Steiner theorem ((Toumanidou & Noailly, 2015), Equation 3). To calculate the upper body loads at UIV+1, a segmental body mass ($SBM_i$) was introduced according to the fixed percentages of the total body mass per vertebral level ((Pearsall et al., 1996), (Vette et al., 2011)).

$$d_{eff} = \left(\frac{Iz_{eff}}{\sum_{i=1}^{Upper} SBM_i}\right)^{-\frac{1}{2}} \tag{3}$$

where $Iz_{eff}$ is the effective moment of inertia at each vertebra level (Toumanidou & Noailly, 2015). BM loads were included in the FE simulation steps for the T10-int and the T11-T3 cases (Table 5).

*Table 5: BM loads applied along the eccentric path for the 1st and 2nd cases.*

| Vertebrae Level | BM loads (N) T10-in | BM loads (N) T11-T3 |
|---|---|---|
| T1 | 14.05 | 17.10 |
| T2 | 15.75 | 18.20 |
| T3 | 16.81 | 19.58 |
| T4 | 17.80 | 20.88 |
| T5 | 18.83 | 22.22 |
| T6 | 19.86 | 23.57 |
| T7 | 20.95 | 24.99 |
| T8 | 22.11 | 26.51 |
| T9 | 23.33 | 28.10 |
| T10 | 24.87 | 30.12 |
| T11 | 26.48 | 32.21 |
| T12 | 28.41 | 34.73 |
| L1 | 30.24 | 37.12 |
| L2 | 32.07 | 39.52 |
| L3 | 33.90 | 41.91 |
| L4 | 35.89 | 44.51 |
| L5 | 37.92 | 47.15 |

### 2.8. Integrated simulation workflow and biomechanical descriptor

The integrated FE simulation workflow is shown in Figure 9. It incorporates an initialization of the intervertebral disc models (Step 1), as these need to swell to capture the effect of osmotic pressurization. Following this initialization, a step of imposed kinematical constraints (Step 2) reflected the mechanical manipulations of the spine in the surgery room. Then, the virtual instrumentation and the application of body mass forces are incorporated, as previously described. The different steps of the workflow are further summarized below.

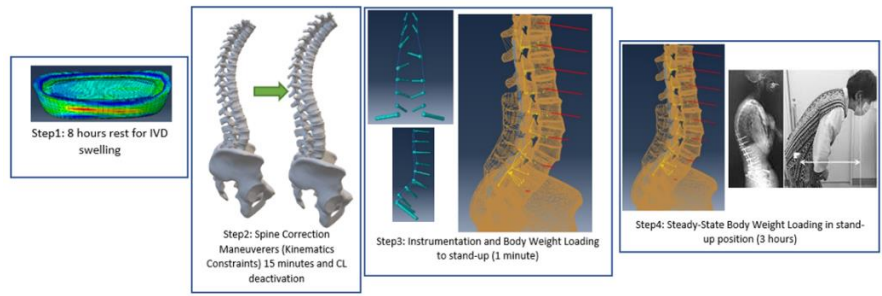

*Figure 9: FE simulation steps*

In Step 1, the IVDs of the T10-int and T11-T3 case models were pre-swelled for 8 hours, without any external load on the modelled thoracolumbar structures, to pressurize the water content in the NP, according to the osmo-poroelastic constitutive model, and pretense the AF fibers (Wills et al., 2016). In Step 2, Capsular Ligaments (CL) were deactivated in the T10-int case model, and kinematic constraints were applied subsequently, to simulate the correction of the spine curvature, until the expected correction and the measured first-op geometrical parameters (Figure 1) were met. Orthopedic surgeons can rotate the vertebrae in the surgery room. Therefore, they can re-shape spine curvature to achieve balanced spine and desired geometrical parameters. Accordingly, the imposed kinematic constraints have been obtained by calculating the rotations among the landmarks of the pre-operated and post-operated vertebras. Rotation was further applied as velocity boundary for vertebra points in step 2 of 1st case. Because the mechanical response of the osteo-ligamentous model is time-dependent, Step 2 used velocity fields to simulate 15 minutes of manipulation, according to the corresponding average time during the surgery at Hospital del Mar, Barcelona. There was no spine correction maneuver (Step 2) for the T11-T3 case.

In simulation Step 3, instrumentation components (screws and rods) were activated for both the T10-int and T11-T3 case models. Contact models between screw and bone were also activated immediately after model activations. BM loads (Table 5) were later applied for each vertebra along the eccentric path in this simulation step. BM loading after surgery was applied in 1 minute, to have the minimum loading time able to secure numerical stability and proper convergence of the poromechanical solver in the modelled IVDs. Eventually, an average time of three hours was considered in the last simulation step (Step 4), to reach steady-state internal loads in the IVD.

In each simulation requiring sequences of activation and de-activation of specific elements, we used the Model Change feature in the Abaqus (SIMULIA, Dassault Industries) software that was used to solve the models. 6 simulations (Figure 7a) for the T10-int case and 7 simulations (Figure 7b) for the T11-T3 were performed, and the T11-T3 modelling workflow was repeated for each of the instrumentation and GAP score model of the virtual cohort. For each simulation, three biomechanical parameters were assessed as FE simulation outputs: 1- maximum principal strain at IVD fibers; 2- screw pull-out force; 3- rod maximum principal stress. Trade-off between fiber strain, screw pull-out, and rod stress could help to select the sub-optimal configurations that best reduce the risk of PJF, for each modelled bone quality reflected by the normal, osteopenia and osteoporotic mechanical properties (Table 4). Further analysis for different GAP scores (7 to 13) and UIV at T10 and T3 was done with the 91 virtual cohort models, to understand better the functional margin of HDR effects versus GAP scores.

## 3. Results
### 3.1. Biomechanical descriptors for the T10-int case

Average fiber maximum principal strain at the UIV/UIV+1 (T10/T9) level for the T10-int case was compared to the control value of 0.1056 (Holzapfel et al., 2005) (Figure 10). To eliminate weak discontinuity artefacts between the IVD and the vertebrae, average maximum principal strain was assessed in AF elements by ignoring the top and bottom elements that share nodes with the elements of the vertebrae. Results showed that HRD, from the real surgical case (scenario 1, Ti rod) to scenarios 2 to 6, led to 23%, 47.76%, 61.47%, 11.53%, and 5.52% fiber strain reductions, respectively (Figure 10). Scenarios 1, 2, 5, and 6 (Ti rod) led to fiber strains larger than the independent control value (0.1056). In contrast, scenarios 3 and 4 led to AF fiber strains lower than the independent control. Moreover, results showed that using Cr-Co instead of Ti rods increased the fiber strains by 20.31% on average for all the scenarios. Among the 6 scenarios, scenario 4 had the highest relative increase (36%) of fiber strain in the case of switching the rod material from Ti to Cr-Co, but only scenario 4 kept the fiber strains below the independent control value in case of using Cr-Co rods (Figure 10). In Scenario 3, the choice of the rod material made the fiber strain cross the control.

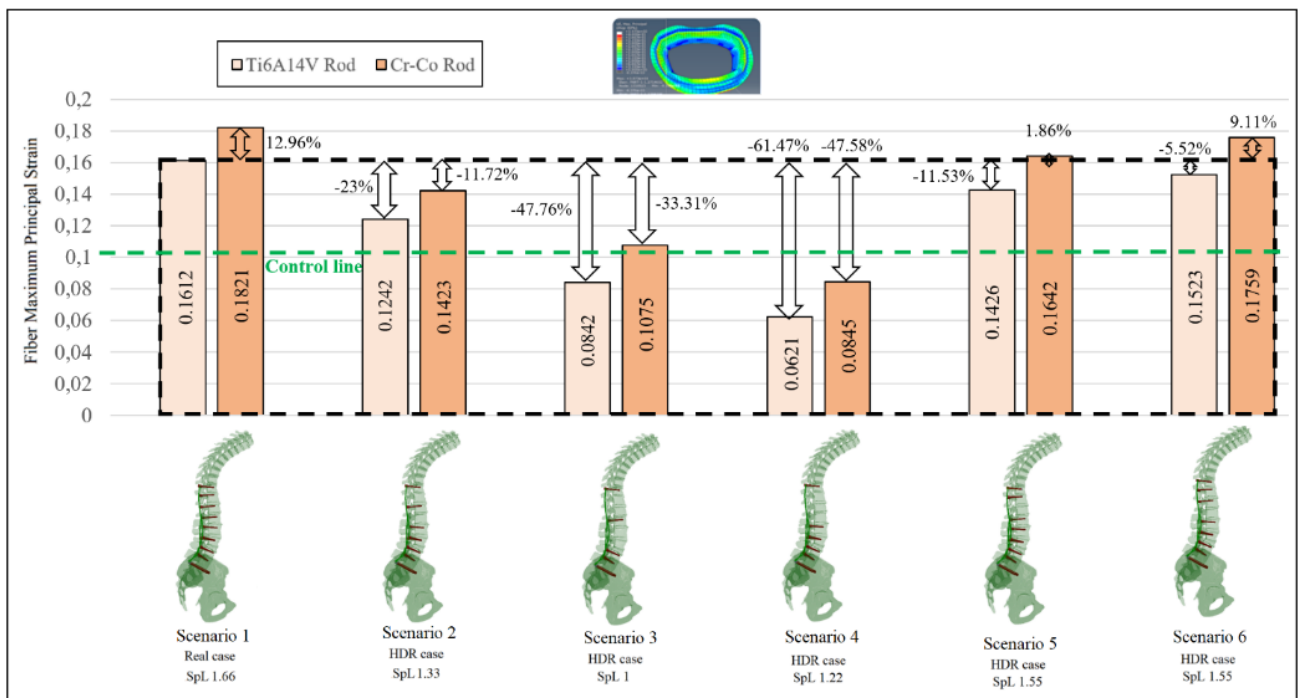

*Figure 10: Maximum principal fiber strain at T10/T9 for 6 scenarios (T10-int case).*

Screw pull-out forces at UIV T10 (10-int case) were assessed for different patterns of fixation with Ti or Cr-Co rods, and with three bone properties, i.e., normal, osteopenia or osteoporotic (Figure 11). HRD, from the real surgical case (scenario 1, Ti rod) to scenarios 2, 3, and 4, 5, and 6 led to 31%, 79%, 101%, 16%, and 7% increase in screw pull-out force, respectively, for the normal bone density (Figure 11). Average screw pull-out forces with Ti rods, for the osteopenia, and for the osteoporotic bones were increased by 30%, and by 55%, respectively, with respect to the normal bone density (Figure 11). Screw pull-out forces in Figure 85 are deemed to be safe, when they are below the three independent threshold values, for each bone quality, calculated by Jendoubi et al., 2018. As such, the 3$^{rd}$ and 4$^{th}$ with Ti rod had a risk of screw pull-out only when the bone was modelled as osteoporotic.

The screw pull-out force in Scenario 2 for the osteoporotic bone and Ti rod was below the osteoporotic independent threshold, but this scenario had a high risk of overstretching the IVD fibers (Figure 10). Moreover, results showed that the change of the rod material, from Ti to Cr-Co, increased the screw pull-out force by 20%, 24.5%, and 30%, in average over the 6 scenarios, for the normal, the osteopenia, and the osteoporotic bone models, respectively. Among the 6 scenarios, scenario 4 had the highest increase rate of screw pull-out force (+36% "normal bone density"; +41% "osteopenia bone"; +46% "osteoporotic bone"), in the case of Ti to Cr-Co change. Scenario 1 had the lowest increase rate of screw pull-out force (+11% "normal bone density", +15% "osteopenia bone", and +20% "osteoporotic bone") from Ti to Cr-Co rods. Although scenario 4 had fiber strains below the control value with Cr-Co rods, the calculated screw pull-out force was below the ultimate values, only for the normal bone density.

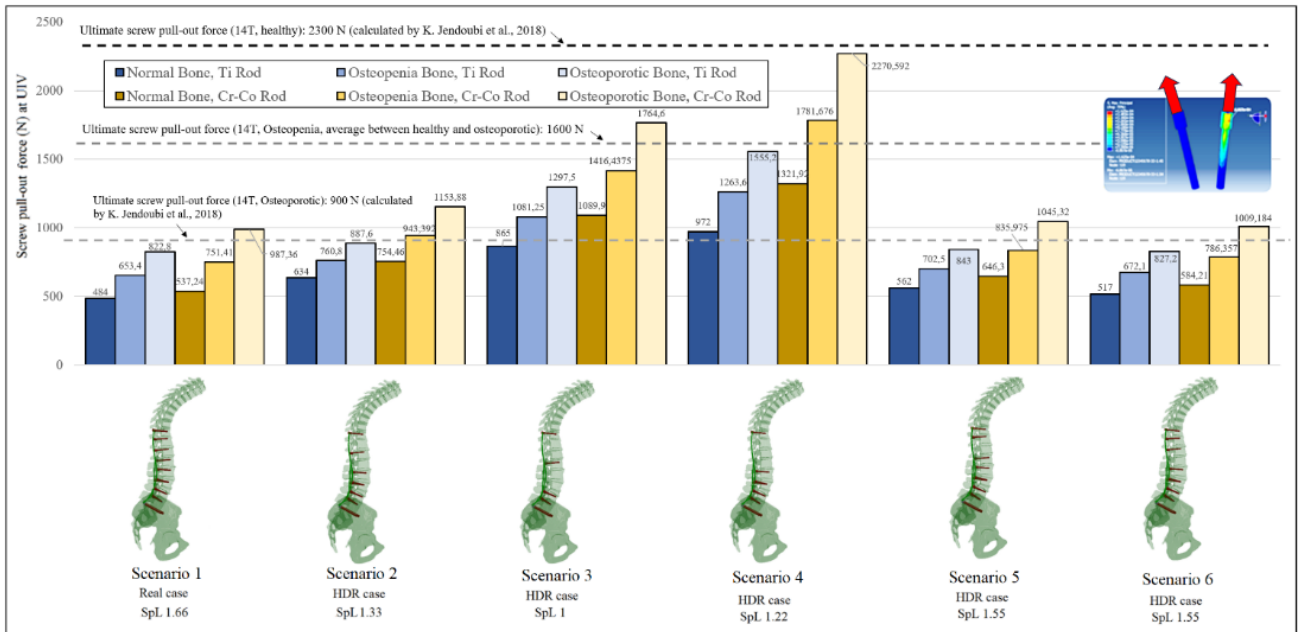

*Figure 11: Pull-out Force (N) at UIV (T10) for 6 scenarios (T10-int case).*

Further analysis showed that none of the calculated rod stresses overcome the Ti or the Cr-Co materials yield stress for the T10-int case.

### 3.2. Biomechanical descriptors for the T11-T3 case

Average and homogenous fiber maximum principal strain at UIV/UIV+1 (T3/T2) level for the T11-T3 case were compared to the independent control value (0.1056, Holzapfel et al., 2005). Results showed that HDR, from the real surgical scenario (scenario 1, Ti rod) to scenarios 2, 3, and 4 led to 41%, 58%, and 70% fiber strain reductions, respectively (Figure 12). Scenarios 2,3,4 (Ti rod) returned fiber strains below the independent control value. The real surgical scenario (scenario 1, Ti rod) and scenarios 6 and 7 all led to fiber strain values larger than the independent control. Furthermore, results showed that changing the rod material, from Ti to Cr-Co rod increased the fiber strains by 19.3% on average for all the scenarios. Among the 6 scenarios, scenarios 4 and 7 had the highest (+36%), and lowest (+9%) rate of fiber strain increase, respectively, because of the use of Cr-Co instrumentation. Only scenario 4 maintained fiber strains below the control value with Cr-Co rod (Figure 12).

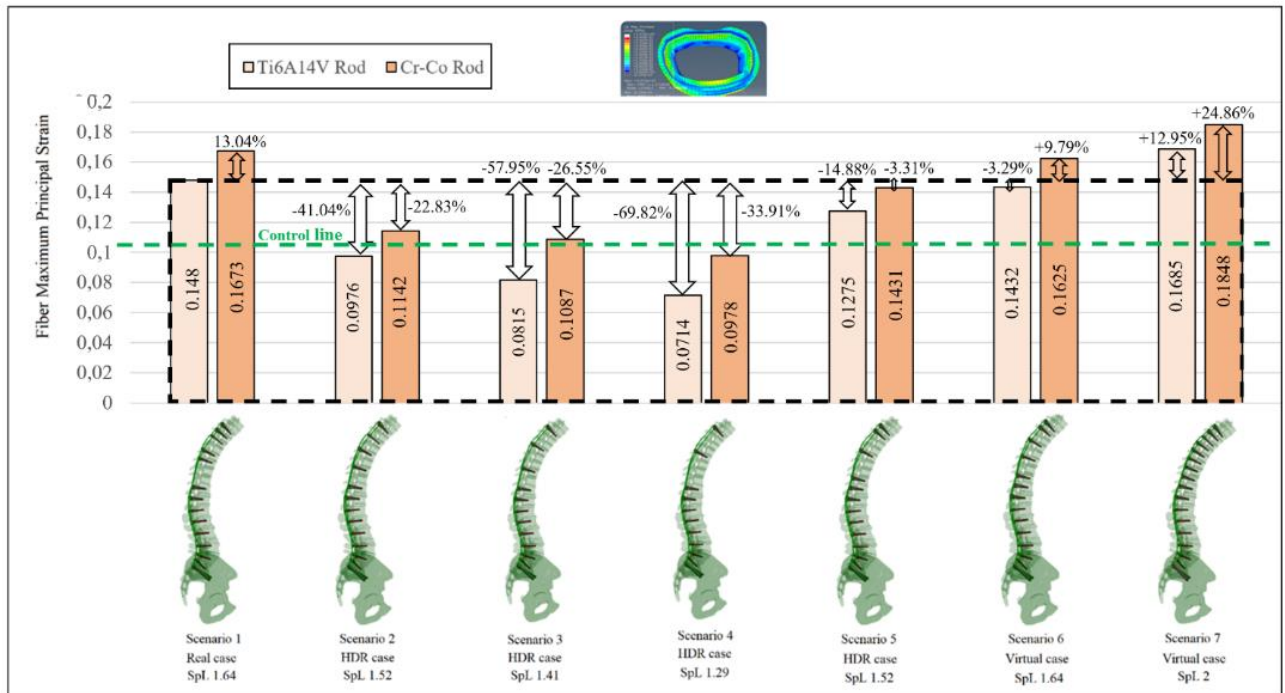

*Figure 12: Maximum principal fiber strain at T3/T2 for 7 scenarios (T11-T3 case).*

Screw pull-out forces at UIV T3 (T11-T3 case) were assessed for different patterns of screw fusion, Ti and Cr-Co rods, and three bone properties of normal, osteopenia, and osteoporotic (Figure 13). Results showed that HDR from the real surgical scenario (scenario 1, Ti rod) to scenarios 2, 3, and 4, 5, and 6 led to 41%, 92%, 119%, 142%, and 4% increase in screw pull-out force for the normal bone density, respectively (Figure 13). Average screw pull-out forces for the osteopenia, and osteoporotic bones in case of Ti rod were increased by 27%, and 56%, respectively, compared to the normal bone density (Figure 13). The 3$^{rd}$ and 4$^{th}$ scenarios in case of Ti rods presented screw pull-out risks when bone was modelled as osteoporotic, according to the corresponding independent ultimate force threshold. Moreover, results showed that change of rod material from Ti to Cr-Co increased the screw pull-out force by 18.7%, 28%, and 23% (in average) for the normal, osteopenia, and osteoporotic bone, respectively. Among the 6 scenarios, scenario 4 had the highest (+36% with normal bone density; +41% with osteopenia bone; +46% with osteoporotic bone) increase rate of screw pull-out force, when rod material properties were changed from Ti to Cr-Co. Although scenario 4 had the fiber strains below the independent control value with Cr-Co rods, screw pull-out force was below the ultimate pull-out force only with normal bone density (Figure 13).

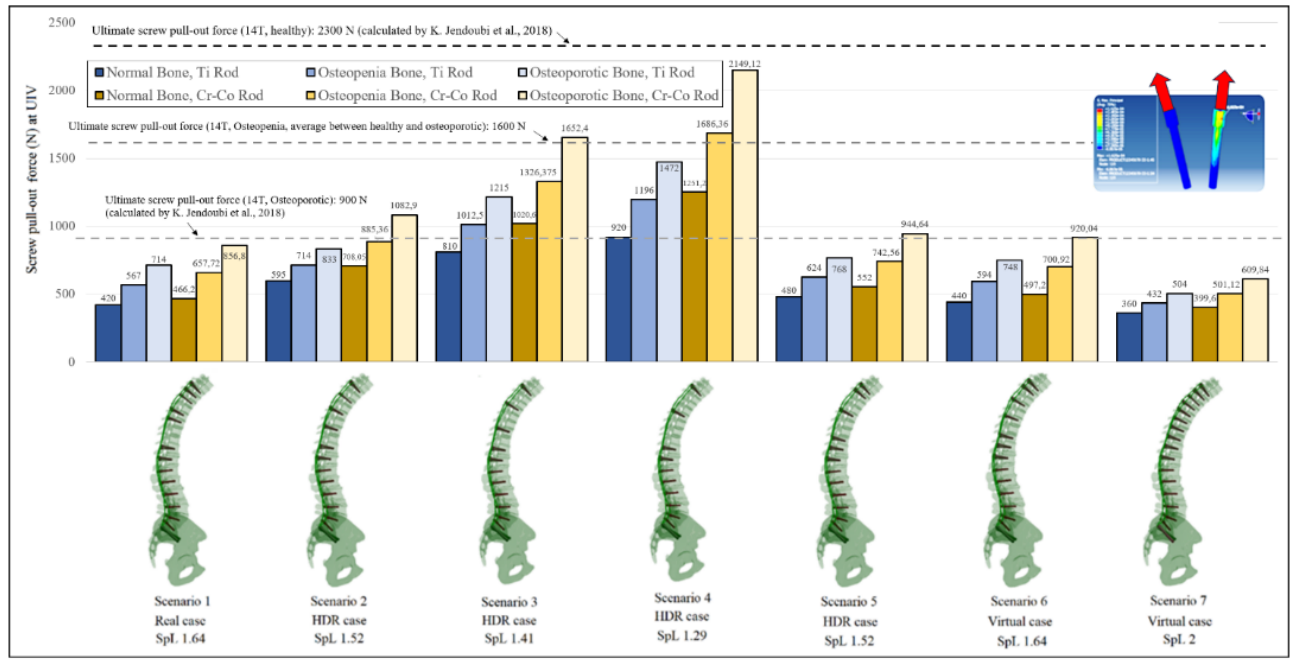

*Figure 13: Pull-out Force (N) on UIV screw for 7 scenarios (T11-T3 case).*

Moreover, results showed that none of the calculated rod stresses overcome the Ti or the Cr-Co materials yield stress for the T11-T3 case.

### 3.3. Risk Trade-off Analyses and Surgery Planning for T10-in & T11-T3 Cases

Based on the results obtained in sections 3.1 and 3.2, a color-coded map (red: risk; green: safe) of the PJF risk associated with fiber maximum principal strain at UIV/UIV+1 and screw pull-out force at UIV is shown in Figure 14, for different bone quality and rod material models. For a specific screw density scenario, a specific rod property is selected first. Then, fiber strain, and screw pull-out force are assessed, for the different bone qualities. If all the quantitative values are in the safe region (green), for a specific scenario and bone quality, the related HDR is recommended for the corresponding bine quality.

| T10-int Case | | | | | | | | |
|---|---|---|---|---|---|---|---|---|
| *Scenarios* | | | 1 | 2 | 3 | 4 | 5 | 6 |
| Ti6Al4V rod | Fiber strain | | 🟥 | 🟥 | 🟩 | 🟩 | 🟥 | 🟥 |
| | Screw pull-out force | Normal bone density | 🟩 | 🟩 | 🟩 | 🟩 | 🟩 | 🟩 |
| | | Osteopenia bone | 🟩 | 🟩 | 🟩 | 🟩 | 🟩 | 🟩 |
| | | Osteoporotic bone | 🟩 | 🟩 | 🟥 | 🟥 | 🟩 | 🟩 |
| Cr-Co rod | Fiber strain | | 🟥 | 🟥 | 🟥 | 🟩 | 🟥 | 🟥 |
| | Screw pull-out force | Normal bone density | 🟩 | 🟩 | 🟩 | 🟩 | 🟩 | 🟩 |
| | | Osteopenia bone | 🟩 | 🟩 | 🟩 | 🟥 | 🟩 | 🟩 |
| | | Osteoporotic bone | 🟥 | 🟥 | 🟥 | 🟥 | 🟥 | 🟥 |
| T11-T3 Case | | | | | | | | |
| *Scenarios* | | | 1 | 2 | 3 | 4 | 5 | 6 | 7 |
| Ti6Al4V rod | Fiber strain | | 🟥 | 🟩 | 🟩 | 🟩 | 🟥 | 🟥 | 🟥 |
| | Screw pull-out force | Normal bone density | 🟩 | 🟩 | 🟩 | 🟩 | 🟩 | 🟩 | 🟩 |
| | | Osteopenia bone | 🟩 | 🟩 | 🟩 | 🟩 | 🟩 | 🟩 | 🟩 |
| | | Osteoporotic bone | 🟩 | 🟩 | 🟥 | 🟥 | 🟩 | 🟩 | 🟩 |
| Cr-Co rod | Fiber strain | | 🟥 | 🟥 | 🟥 | 🟩 | 🟥 | 🟥 | 🟥 |
| | Screw pull-out force | Normal bone density | 🟩 | 🟩 | 🟩 | 🟩 | 🟩 | 🟩 | 🟩 |
| | | Osteopenia bone | 🟩 | 🟩 | 🟩 | 🟥 | 🟩 | 🟩 | 🟩 |
| | | Osteoporotic bone | 🟩 | 🟥 | 🟥 | 🟥 | 🟥 | 🟥 | 🟩 |

*Figure 14: Trade-off analyses between fiber strain, screw pull-out force, and rod stress (green means allowed, red means rejected).*

Results showed that the calculated screw pull out forces appear to stand for a risk, neither for the T10-int or for the T11-T3 case, always when the bone is not osteoporotic and Ti rods are used. In contrast, the use of Cr-Co rods might represent a risk in scenarios 4, when the bone is either osteoporotic or suffers from osteopenia. Moreover, considering fiber strains in a T10-in surgery case, the choice of Ti rods would allow HDR scenario 3 and 4 to be safe in a non-osteoporotic patient. However, no safe configuration at all could be found is the T10-in suffers from osteoporosis. Should Cr-Co rods be chosen, only case 4 would be safe if the T10-int patient has normal bone density properties. Furthermore, by considering fiber strains in the T11-T3 case and Ti rods, HDR scenario 1 would be fully safe, while the safety of Scenarios 3 and 4 would be limited to non-osteoporotic bone quality. Should Cr-Co rods be chosen, only HRD scenario 4 would be safe, only if the T11-T3 patient has a normal bone quality.

### 3.4. HDR vs. GAP analysis

91 FE simulations were done to test different screw density configurations in our virtual population with different GAP scores. Three biomechanical descriptors were assessed: fiber strains, screw pull-out force, and rod stress, and a risk trade-off analysis as in 3.3 was done. The results of which are reported in Table 6, in terms of allowed SpL.

*Table 6: Allowed configurations in different GAP scores.*

| UIV | GAP scores | Rod properties | Allowed SpL(s) (trade-off between three biomechanical descriptors) | Safe bone quality (screw pull-out force assessment) |
|---|---|---|---|---|
| T10 | 7 | Ti6Al4V | 1.5, 1.75 [screw at T12 removed], 1.75 [screw at L1 removed] | normal, osteopenia, and osteoporotic |
| | | | 1, 1.25 | normal, and osteopenia |
| | | Cr-Co | 1.5 | normal, osteopenia, and osteoporotic |
| | | | 1, 1.25 | normal, and osteopenia |
| | 8 | Ti6Al4V | 1.5, 1.75 [screw at T12 removed] | normal, osteopenia, and osteoporotic |
| | | | 1, 1.25 | normal, and osteopenia |
| | | Cr-Co | 1.5 | normal, osteopenia, and osteoporotic |
| | | | 1, 1.25 | normal, and osteopenia |
| | 9 | Ti6Al4V | 1.5 | normal, osteopenia, and osteoporotic |
| | | | 1, 1.25 | normal, and osteopenia |
| | | Cr-Co | 1, 1.25 | normal, and osteopenia |
| | 10 | Ti6Al4V | 1, 1.25 | normal, and osteopenia |
| | | Cr-Co | 1.25 | normal |
| | 11 | Ti6Al4V | 1.25 | normal |
| | | Cr-Co | --- | --- |
| | 12 | Ti6Al4V | --- | --- |
| | | Cr-Co | --- | --- |
| | 13 | Ti6Al4V | --- | --- |
| | | Cr-Co | --- | --- |
| T3 | 7 | Ti6Al4V | 1.52 (1), 1.52 (2), 1.64 (2), 1.41, 1.64 (1) | normal, osteopenia, and osteoporotic |
| | | | 1.29 | normal, and osteopenia |
| | | Cr-Co | 1.29, 1.41 | normal, and osteopenia |
| | | | 1.52 (1), 1.52 (2), 1.64 (2), 1.64 (1) | normal, osteopenia, and osteoporotic |
| | 8 | Ti6Al4V | 1.52 (1), 1.52 (2), 1.64 (2), 1.41, 1.64 (1) | normal, osteopenia, and osteoporotic |
| | | | 1.29 | normal, and osteopenia |
| | | Cr-Co | 1.29, 1.41 | normal, and osteopenia |
| | | | 1.52 (1), 1.52 (2), 1.64 (2) | normal, osteopenia, and osteoporotic |
| | 9 | Ti6Al4V | 1.52 (1), 1.52 (2), 1.64 (2), 1.41 | normal, osteopenia, and osteoporotic |
| | | | 1.29 | normal, and osteopenia |
| | | Cr-Co | 1.29, 1.41 | normal, and osteopenia |
| | | | 1.52 (1), 152 (2) | normal, osteopenia, and osteoporotic |
| | 10 | Ti6Al4V | 1.52 (1), 1.52 (2) | normal, osteopenia, and osteoporotic |
| | | | 1.29, 1.41 | normal, and osteopenia |
| | | Cr-Co | 1.29, 1.41 | normal, and osteopenia |
| | | | 1.52 (1) | normal, osteopenia, and osteoporotic |
| | 11 | Ti6Al4V | 1.52 (1) | normal, osteopenia, and osteoporotic |
| | | | 1.29, 1.41 | normal, and osteopenia |
| | | Cr-Co | 1.29 | normal |
| | 12 | Ti6Al4V | 1.29, 1.41 | normal |
| | | Cr-Co | --- | --- |
| | 13 | Ti6Al4V | --- | --- |
| | | Cr-Co | --- | --- |

Results in Table 6 showed GAP 12 and 13 in an UIV T10 virtual patient could not lead to any safe HDR scenario in terms of PJF, either with Ti or Cr-Co rods. HDR in a UIV T10 virtual patient with GAP 11 could not de-risk the patient if Cr-Co rods are used. Should UIV be T3, GAP 13 virtual patients would not be able to benefit from any HDR strategy, independently of the material of the rods. In contrast, the use of Ti rods might allow HDR to de-risk GAP 12 patients with UIV T3. These results confirm that the possible success of HDR strategies is both UIV- and GAP-dependent.

## 4. Discussion

Screw density reduction (HDR) in spine surgery aims to reduce the rigidity of the overall spinal fixation construct, to limit local stress concentrations while maintaining the necessary mechanical stability of the operated spine (Shin, 2013). Several studies have focused on the effect of HDR on the outcomes of scoliotic spine surgery (Shen et al., 2017; Chotigavanichaya et al., 2023; Sariyilmaz et al., 2018; Larson et al., 2014; Liu et al., 2015; Luo et al., 2017). But only a few studies showed the correlation between pedicle screw density and spine sagittal alignment to treat pure sagittal deformities. Behrbalk et al. showed that the average postoperative kyphosis correction was similar in both high and low screw density groups, in the case of corrective surgery of Scheuermann kyphosis (SK) (Behrbalk et al., 2014). The Authors further asserted that low screw densities led to less post-operative complications compared to high screw densities in posterior-only correction of SK (Behrbalk et al., 2014). Durand et al, found that patients with SpL lower than 1.8 exhibited reduced odds of PJF (Durand et al., 2022). In contrast, the real clinical cases, T10-int and T11-T3, presented hereby suffer from PJF with SpL values of 1.66, and 1.64, respectively, and simulations suggested that SpL should be lowered at least to 1.41 and 1.52, for these T10-int and T11-T3 cases, respectively. This can be explained by the fact that PJF is a local event at the cranial end of the instrumentation, whereas SpL is a global metrics over the entire treated section of the spine. Hence, SpL lacks information about the distribution of the screws, and about the interplay with the GAP score. To address this issue, FE simulations and biomechanical descriptors in this study were able to identify the best distribution patterns of screws which prevent PJF.

On the one hand, a systematic analysis of the literature, Hashimoto et al. showed that excessive motions of adjacent segments in spinal fusion surgeries are likely lead to adjacent segment degeneration or PJK (Hashimoto et al., 2019). On the other hand, simulations with a T12-sacrum FE model by Park et al. have revealed increased stresses in the intervertebral discs and vertebra proximal to the fused levels (W. M. Park et al., 2015). Jang et al. also studied PJF risk factors at the early stage of PJF (no instrumental problems, small PJK angles, or neurologically intact) (Jang et al., 2021). He found that pre-operatively unhealthy discs appear to collapse more easily at the early stage due to a lower capability of absorbing the axial loading (Jang et al., 2021). The anisotropic fiber reinforcement of the annulus fibrosus in the IVD allows the organ to withstand the complex patterns that develop in its circumferential regions, which impedes excessive motions, as nicely shown in several FE studies (Noailly et al., 2011; O'Connell et al., 2011) Moreover, according to the results of scenario 1 in the T10-in and T11-T3 personalized model simulations, fiber strain was decreased mostly by 61.47%, leading to introduce fiber strain as a biomechanical descriptor which altered remarkably when hardware density decreased. Furthermore, fiber strain appeared as the most limiting biomechanics descriptor to assess the potential of HDR, and it pointed out the likely importance of this descriptor to evaluate personalized risks of PJF. Accordingly, IVD fiber strain was chosen hereby as a potential PJF biomechanical descriptor at UIV/UIV+1, and its predicted strains were compared to an independent control strain value of 0.1056 for annulus fibrosus lamellar damage (Holzapfel et al., 2005).

HDR effect on fiber strain was evaluated in 2 clinical cases of T10-int and T11-T3 cases. Results showed that HRD, from the real surgical case of T10-int (scenario 1, Ti rod) to scenarios 2 to 6, led to 23%, 47.76%, 61.47%, 11.53%, and 5.52% fiber strain reductions, respectively. It uncovered this fact that SpL reduction by 39% (1.66 to 1) could decrease fiber strain by 47.76%. These results are in accordance with Shin et al. who found that reducing screw density effectively decreases the rigidity of the construct, thereby increasing flexibility and avoids the stress concentration between fused and mobile spine (Shin, 2013). In contrary, results here showed that HRD, from the real surgical case of T10-int (scenario 1, Ti rod) to scenarios 2, 3, and 4, 5, and 6 led to 31%, 79%, 101%, 16%, and 7% increase in screw pull-out force, respectively, for the normal bone density. These results are also in accordance with Doodkorte et al. who found that screw density reduction led to increased screw pull-out forces (Doodkorte, Vercoulen, et al., 2021). Further results also showed that average screw pull-out forces with Ti rods, for the osteopenia, and for the osteoporotic bones were increased by 30%, and by 55%, respectively, with respect to the normal bone density. These results were also aligned with Jendoubi et al. who found that screw pull-out force varied depending on the bone quality (Jendoubi et al., 2018). Results for Ti and Cr-Co rod demonstrated that the change of the rod material, from Ti to Cr-Co, increased the screw pull-out force by 20%, 24.5%, and 30%, in average over the 6 scenarios of case T10-int, for the healthy, the osteopenia, and the osteoporotic bone models, respectively. Han et al. also showed that Ti rods had lower incidence of PJK and increasing the rod stiffness using cobalt chrome (Cr-Co) could prevent rod breakage but adversely increased the occurrence and the time of PJK (Han et al., 2017). As per the risk of rod failure, rod stresses were compared to the rod material yield stress, and simulations suggested that the tested HDR would not mechanically harm the rods. However, further fatigue analyses are necessary to determine the number of fatigue cycles required for the rod to fracture, also considering the effect of body fluid corrosive environments.

Results showed that HRD, from the real surgical case of T11-T3 (scenario 1, Ti rod) to scenarios 2, 3, and 4, led to 41%, 58%, and 70% fiber strain reductions, respectively. In contrary, results showed that HRD, from the real surgical case of T11-T3 (scenario 1, Ti rod) to scenarios 2, 3, and 4, 5, and 6 led to 41%, 92%, 119%, 142%, and 4% increase in screw pull-out force for the normal bone density, respectively. Further results also showed that average screw pull-out forces for the osteopenia, and osteoporotic bones in case of T11-T3, and Ti rod were increased by 27%, and 56%, respectively, compared to the normal bone density. Moreover, results showed that change of rod material from Ti to Cr-Co increased the screw pull-out force by 18.7%, 28%, and 23% (in average, T11-T3 case) for the healthy, osteopenia, and osteoporotic bone, respectively.

Results also suggest that if an orthopedic surgeon decides to implant the instrumentations from S1 to T10 and cannot restore the normal curvature of the spine (GAP 10, T10-int case), screw distributions of 2 and 3 levels left (scenarios 3 and 4, T10-in case) might avoid early PJF in case of healthy and osteopenia bone and using Ti rod. However, if the bone is osteoporotic, none of the HDR scenarios would be safe, neither with Ti nor with Cr-Co rods. Hence, additional PJF prevention techniques like cement augmentation screw, or laminar hooks might be recommended. Moreover, orthopedic surgeons might only be able to use Cr-Co rods in the scenario 4, in case of normal bone density because of increased risk of screw pull-out with osteopenia and osteoporotic bone. These results are in accordance with Han et al. who showed that Cr-Co rod increased PJF risk (Han et al., 2017). Moreover, if a patient had PJF at T11 with severe GAP 11 (T11-T3 case), leading to extend hardware to T3, following conditions can avoid PJF: screw distribution of one level left (scenario 2, case 2) accompanying Ti rod. According to our simulation results, clinicians might avoid 2$^{nd}$ PJF in the revision surgery using 2 and 3 levels HDR (scenarios 3 and 4, T11-T3 case) and Ti rod, in case of healthy and osteopenia bone. If the bone is osteoporotic, none of the scenarios would be safe, and additional PJF prevention techniques would be needed. Furthermore, orthopedic surgeons could use Cr-Co rod in scenario 4 only in case of normal bone density.

The first aim of spine sagittal surgery is to restore the spine alignment as much as possible. However, because of age-related problems, lumbar osteotomy and excessive spine correction maneuver are avoided. Virtual cohort study showed that HDR effects are GAP-dependent, and the functional margin of HDR strategies are narrowed down in high GAP scores. For example, HDR in GAP 12 and 13 might not reduce the risk of PJF either with Ti or with Cr-Co rods in UIV T10. HDR in GAP 11 might not increase safety with Cr-Co rods in UIV T10, either. Moreover, HDR in GAP 13 wouldn't be efficient independently on the rod material, in UIV T3. HDR in GAP 12 might be useful, only with Ti rods in UIV T3. In contrast, GAP 7 allows a broad range of functional HDR scenarios. These results obtained with the virtual cohort are in accordance with clinical data reported in Table 3. Moreover, some results in Table 6 were validated with the clinical data presented in Table 3. For example, a clinical case with UIV T10, GAP 8, and an HDR of SpL 1.5 (Table 3) was a control case after a period of 22 months. On the other hand, the results in Table 6 also confirmed the effectiveness of both Ti and Cr-Co rods in preventing PJF in patients with healthy, osteopenia, or osteoporotic bone.

In addition to HDR, biomechanical responses of other topping-off techniques like posteriorly anchored polyethylene tethers, e.g., calculated through patient-specific FE analyses (Buell et al., 2019), may help orthopedic surgeons to manage complex surgery plan by integrating different PJF prevention techniques. Whether the rib cage resistance, cement augmentation screws, and muscle tension and fatigue would modify the assessment of the risk of PJF through the hereby studied biomechanical descriptors remains unclear, and probably require further investigations involving computational biomechanical models (Toumanidou & Noailly, 2015; Ignasiak et al., 2018). The present biomechanical study only considers the effect of the anterior body weight distribution and stands for a worst-case scenario that was shown, however, very effective to discriminate PJF in a retrospective case-control clinical study (Rasouligandomani et al., 2023) .

Other limitations for this study were lack of basic data, i.e., weight or BMI of case studies, which forced us to exploit the correlation models to achieve BMI values. To address these issues, O'Neill (O'Neill et al., 2018) model was used, and end results were verified through the alternative BMI Bozeman (Bozeman et al., 2012) model (more details presented in Rasouligandomani et al., 2023). Larger clinical cohort would be needed to validate screw density results and build proper statistics. Sarcopenia and osteoporosis information for clinical cases were not available. Because of this uncertainty, a parametric analysis was performed, to check the screw pull-out resistance in different bone properties.

## 5. Conclusion

A unique campaign of personalized FE simulations of real clinical cases and of virtual surgical scenarios was done, to cope with ASD, in the case of suboptimal capacity of sagittal balance correction. As a conclusion, HDR could aid spine orthopedic surgeons to avoid PJF even with severe sagittal deformity, depending on the screw distribution at proximal levels of fixation, bone material property, and rod material property. This study showed that HDR reduces the chronic load of the upper IVD, and it might help to decrease risk of PJF in case of suboptimal surgical decision. Trade-off analyses between fiber strains and screw pull-out force in different bone properties, and with different rod materials exploited the allowable screw density configurations which avoid PJF. Spine orthopedic surgeons can make the post-op GAP score estimation to guide their selection of configurations which prevent PJF. HDR technique also avoids excessive spine surgeries, hospitalization stay, and blood loss, decreases mortality rate, and minimizes the surgery costs. Effects of paravertebral muscles and dynamic loads, other mechanical complications like rod breakage, distal failure etc., testing more clinical cases, and uncertainty analysis to assess the error propagation rate should be topic of future works.

## 6. Acknowledgments

Funds received from DTIC-UPF, IMIM, IRCCS Istituto Ortopedico Galeazzi, the Spanish Government (CEX2021-001195-M), and the European Commission and the European Research Council (ERC-2021-CoG-O-Health-101044828). Mobility grant also received from European Society of Biomechanics (ESB), extraordinary call 2020.

Hashimoto, K., Aizawa, T., Kanno, H., & Itoi, E. (2019). Adjacent segment degeneration after fusion spinal surgery—a systematic review. *International Orthopaedics*, *43*(4), 987–993. https://doi.org/10.1007/s00264-018-4241-z

Hirose, O. (2021). A Bayesian Formulation of Coherent Point Drift. *IEEE Transactions on Pattern Analysis and Machine Intelligence*, *43*(7), 2269–2286. https://doi.org/10.1109/TPAMI.2020.2971687

Holzapfel, G. A., Schulze-Bauer, C. A. J., Feigl, G., & Regitnig, P. (2005). Single lamellar mechanics of the human lumbar anulus fibrosus. *Biomechanics and Modeling in Mechanobiology*, *3*(3), 125–140. https://doi.org/10.1007/s10237-004-0053-8

Ignasiak, D., Valenzuela, W., Reyes, M., & Ferguson, S. J. (2018). The effect of muscle ageing and sarcopenia on spinal segmental loads. *European Spine Journal*, *27*(10), 2650–2659. https://doi.org/10.1007/s00586-018-5729-3

Jacobs, E., Roth, A. K., Arts, J. J., van Rhijn, L. W., & Willems, P. C. (2017). Reduction of intradiscal pressure by the use of polycarbonate-urethane rods as compared to titanium rods in posterior thoracolumbar spinal fixation. *Journal of Materials Science: Materials in Medicine*, *28*(10), 148. https://doi.org/10.1007/s10856-017-5953-0

Jang, H. J., Park, J. Y., Kuh, S. U., Chin, D. K., Kim, K. S., Cho, Y. E., Hahn, B. S., & Kim, K. H. (2021). The Fate of Proximal Junctional Vertebral Fractures after Long-Segment Spinal Fixation : Are There Predictable Radiologic Characteristics for Revision surgery? *Journal of Korean Neurosurgical Society*, *64*(3), 437–446. https://doi.org/10.3340/jkns.2020.0236

Jendoubi, K., Khadri, Y., Bendjaballah, M., & Slimane, N. (2018). Effects of the Insertion Type and Depth on the Pedicle Screw Pullout Strength: A Finite Element Study. *Applied Bionics and Biomechanics*, *2018*, 1–9. https://doi.org/10.1155/2018/1460195

Jensen, K. S., Mosekilde, L., & Mosekilde, L. (1990). A model of vertebral trabecular bone architecture and its mechanical properties. *Bone*, *11*(6), 417–423. https://doi.org/10.1016/8756-3282(90)90137-N

Jolliffe, I. T., & Cadima, J. (2016). Principal component analysis: a review and recent developments. *Philosophical Transactions of the Royal Society A: Mathematical, Physical and Engineering Sciences*, *374*(2065), 20150202. https://doi.org/10.1098/rsta.2015.0202

Kasra, M., Vanin, C. M., MacLusky, N. J., Casper, R. F., & Grynpas, M. D. (1997). Effects of different estrogen and progestin regimens on the mechanical properties of rat femur. *Journal of Orthopaedic Research*, *15*(1), 118–123. https://doi.org/10.1002/jor.1100150117

Kim, K.-T., Park, K.-J., & Lee, J.-H. (2009). Osteotomy of the Spine to Correct the Spinal Deformity. *Asian Spine Journal*, *3*(2), 113. https://doi.org/10.4184/asj.2009.3.2.113

Larson, A. N., Polly, D. W., Diamond, B., Ledonio, C., Richards, B. S., Emans, J. B., Sucato, D. J., & Johnston, C. E. (2014). Does Higher Anchor Density Result in Increased Curve Correction and Improved Clinical Outcomes in Adolescent Idiopathic Scoliosis? *Spine*, *39*(7), 571–578. https://doi.org/10.1097/BRS.0000000000000204

Park, W. M., Choi, D. K., Kim, K., Kim, Y. J., & Kim, Y. H. (2015). Biomechanical effects of fusion levels on the risk of proximal junctional failure and kyphosis in lumbar spinal fusion surgery. *Clinical Biomechanics*, *30*(10), 1162–1169. https://doi.org/10.1016/j.clinbiomech.2015.08.009

Pearsall, D. J., Reid, J. G., & Livingston, L. A. (1996). Segmental inertial parameters of the human trunk as determined from computed tomography. *Annals of Biomedical Engineering*, *24*(2), 198–210. https://doi.org/10.1007/BF02667349

Rasouligandomani, M., del Arco, A., Pellis´e, F., Gonz´alez Ballester, M. Á., Galbusera, F., & Noailly, J. (2023). Proximal Junction Failure in Spine Surgery: Integrating Geometrical and Biomechanical Global Descriptors Improves GAP Score-Based Assessment. *Spine*. https://doi.org/10.1097/BRS.0000000000004630

Sariyilmaz, K., Ozkunt, O., Karademir, G., Gemalmaz, H. C., Dikici, F., & Domanic, U. (2018). Does pedicle screw density matter in Lenke type 5 adolescent idiopathic scoliosis? *Medicine*, *97*(2), e9581. https://doi.org/10.1097/MD.0000000000009581

Shen, M., Jiang, H., Luo, M., Wang, W., Li, N., Wang, L., & Xia, L. (2017). Comparison of low density and high density pedicle screw instrumentation in Lenke 1 adolescent idiopathic scoliosis. *BMC Musculoskeletal Disorders*, *18*(1), 336. https://doi.org/10.1186/s12891-017-1695-x

Shin, B. J. , et al. (2013). The biomechanical properties of a pedicle subtraction osteotomy can be predicted and improved with finite element modeling. *Spine*, *38*(19), E1189–E1195.

Shirazi-Adl, A., Ahmed, A. M., & Shrivastava, S. C. (1986). A finite element study of a lumbar motion segment subjected to pure sagittal plane moments. *Journal of Biomechanics*, *19*(4), 331–350. https://doi.org/10.1016/0021-9290(86)90009-6

Toumanidou, T., & Noailly, J. (2015). Musculoskeletal Modeling of the Lumbar Spine to Explore Functional Interactions between Back Muscle Loads and Intervertebral Disk Multiphysics. *Frontiers in Bioengineering and Biotechnology*, *3*. https://doi.org/10.3389/fbioe.2015.00111

Varghese, V., Saravana Kumar, G., & Krishnan, V. (2017). Effect of various factors on pull out strength of pedicle screw in normal and osteoporotic cancellous bone models. *Medical Engineering & Physics*, *40*, 28–38. https://doi.org/10.1016/j.medengphy.2016.11.012

Vette, A. H., Yoshida, T., Thrasher, T. A., Masani, K., & Popovic, M. R. (2011). A complete, non-lumped, and verifiable set of upper body segment parameters for three-dimensional dynamic modeling. *Medical Engineering & Physics*, *33*(1), 70–79. https://doi.org/10.1016/j.medengphy.2010.09.008

Wills, C. R., Malandrino, A., van Rijsbergen, MM., Lacroix, D., Ito, K., & Noailly, J. (2016). Simulating the sensitivity of cell nutritive environment to composition changes within the intervertebral disc. *Journal of the Mechanics and Physics of Solids*, *90*, 108–123. https://doi.org/10.1016/j.jmps.2016.02.003

Xu, M., Yang, J., Lieberman, I. H., & Haddas, R. (2019). Finite element method-based study of pedicle screw–bone connection in pullout test and physiological spinal loads. *Medical Engineering & Physics*, *67*, 11–21. https://doi.org/10.1016/j.medengphy.2019.03.004
28

Yagi, M., Nakahira, Y., Watanabe, K., Nakamura, M., Matsumoto, M., & Iwamoto, M. (2020). The effect of posterior tethers on the biomechanics of proximal junctional kyphosis: The whole human finite element model analysis. *Scientific Reports*, *10*(1), 3433. https://doi.org/10.1038/s41598-020-59179-w

Yilgor, C., Sogunmez, N., Boissiere, L., Yavuz, Y., Obeid, I., Kleinstück, F., Pérez-Grueso, F. J. S., Acaroglu, E., Haddad, S., Mannion, A. F., Pellise, F., & Alanay, A. (2017). Global Alignment and Proportion (GAP) Score. *Journal of Bone and Joint Surgery*, *99*(19), 1661–1672. https://doi.org/10.2106/JBJS.16.01594

Yin, S., Njai, R., Barker, L., Siegel, P. Z., & Liao, Y. (2016). Summarizing health-related quality of life (HRQOL): development and testing of a one-factor model. *Population Health Metrics*, *14*(1), 22. https://doi.org/10.1186/s12963-016-0091-3

Zhu, W.-Y., Zang, L., Li, J., Guan, L., & Hai, Y. (2019). A biomechanical study on proximal junctional kyphosis following long-segment posterior spinal fusion. *Brazilian Journal of Medical and Biological Research*, *52*(5). https://doi.org/10.1590/1414-431x20197748
29